\documentclass[final,5p,times,twocolumn]{elsarticle}

\usepackage{amssymb}
\usepackage{amsmath}
\usepackage{amsthm}

 \biboptions{comma,sort&compress}

\usepackage{xcolor}
\usepackage[colorlinks=true]{hyperref}
\usepackage{listings}
\usepackage{ragged2e}
\usepackage{cleveref}
\usepackage{slashed}
\usepackage{caption}
\usepackage{subcaption}

\usepackage{booktabs}
\usepackage{multirow}
\usepackage{array}
\usepackage{tabularx}
\usepackage{makecell}
\usepackage{soul}
\usepackage{placeins}

\newcommand{\FORM}{\textsc{Form}}
\newcommand{\OPITER}{\textsc{OPITeR}}

\newcommand{\ixset}[2]{{#1_1\dots #1_{#2}}}

\DeclareMathOperator{\sgn}{sgn}
\DeclareMathOperator{\tr}{tr}

\newcommand{\dd}{\mathrm{d}}

\newcommand{\review}[1]{#1}

\usepackage{listings}
\usepackage{courier}
\definecolor{backcolour}{RGB}{255,230,240}

\lstdefinestyle{greybox}{
  backgroundcolor=\color{backcolour},
  basicstyle=\ttfamily\small,
  framexleftmargin=5pt, 
  framexrightmargin=5pt,
  framextopmargin=2pt,
  framexbottommargin=2pt,
  frame=lines,
}
\newcommand{\codeparaheading}[1]{\colorbox{backcolour}{\lstinline|#1|}}

\newcounter{bla}

\journal{Computer Physics Communications}

\begin{document}

\begin{frontmatter}



\title{\OPITER: A program for tensor reduction of multi-loop Feynman Integrals}


\author[a]{Jae Goode}
\author[a]{Franz Herzog\corref{author}}
\author[a]{Sam Teale}

\cortext[author] {Franz Herzog.\\\textit{E-mail address:} fherzog@ed.ac.uk}
\address[a]{Higgs Centre for Theoretical Physics, School of Physics and Astronomy, The University
of Edinburgh, Edinburgh, EH9 3FD, Scotland, UK}

\begin{abstract}




We present \OPITER, a \FORM{} program for the reduction of multi-loop tensor Feynman integrals. The program can handle tensors, including  spinor indices, with rank of up to 20 and can deal with up to 8 independent external momenta. The reduction occurs in $D$ dimensions compatible with conventional dimensional regularization. The program is able to manifest symmetries of the integrand in the tensor reduced form.

\end{abstract}


\end{frontmatter}


\noindent {\bf PROGRAM SUMMARY}

\begin{small}
\noindent
{\em Program Title: }  \OPITER                            \\
{\em Developer's repository link: }
\href{https://bitbucket.org/jaegoode/opiter}{bitbucket.org/jaegoode/opiter}  \\
{\em Licensing provisions:} GPLv3 \\
{\em Programming language:} \FORM{} [1]   \\
{\em Nature of problem:} Tensor Feynman integrals, including both Lorentz and spinor indices, require reduction to scalar integrals. At high ranks, especially for high loops and many external momenta, this problem leads to large, dense systems of equations with standard approaches.\\
{\em Solution method:}
The orbit partition approach [2] leads to a combinatorial solution for arbitrary tensor Feynman integrals, naturally implemented with \FORM{}'s built-in commands.\\
{\em Additional comments including restrictions and unusual features:}\\
The tensor rank must be less than 22. The number of independent external momenta must be less than 9.

\end{small}

\section{Introduction}

Evaluating loop integrals is vital to calculations in perturbative quantum field theory. The standard approach to such integrals is the reduction to a set of master integrals (MIs) which are a set of linearly independent (Lorentz) scalar integrals. Before the reduction to MIs -- conventionally through integration-by-parts identities (IBPs) \cite{Tkachov:1981wb,Chetyrkin:1981qh,Laporta:2000dsw} -- a tensor integral must first be reduced to scalar ones.

There are many algorithms and methods to perform tensor reduction. Methods based on  Passarino-Veltmann reduction \cite{Passarino:1978jh} rely on a general ansatz in terms of  all possible Lorentz structures composed of metric tensors and momenta in the problem. For high tensor rank, solving for the ansatz unknowns requires the solving of a large, dense system of equations.
At one loop a variety of elegant solutions have been developed \cite{Ezawa:1990dh,Devaraj:1997es,Denner:2005nn,Binoth:2008uq,Diakonidis:2008ij,Diakonidis:2009fx, Fleischer:2010sq,Fleischer:2011bi,Fleischer:2011zz,Fleischer:2011zx,Fleischer:2011hc,Fleischer:2011nt}. The problem may also be circumvented with unitarity-based methods \cite{Ossola:2006us,Forde:2007mi,Giele:2008ve,Berger:2008sj} or contraction with \review{auxiliary vectors \cite{Feng:2021enk,Hu:2021nia,Feng:2022hyg,Feng:2022uqp,Feng:2022rfz,Feng:2022iuc,Jin:2022zcj,Hu:2024}}. At higher loops more techniques have been employed: the projector-based approach  applied to on-shell amplitudes \cite{Binoth:2002xg,Chen:2019wyb}, and its extension in the 't Hooft-Veltmann scheme \cite{Peraro:2019cjj,Peraro:2020sfm, Gehrmann:2022vuk,Gehrmann:2023zpz}; unitarity-based approaches \cite{Ita:2015tya,Abreu:2017hqn,Badger:2017jhb}; projectors based on  differential operators \cite{Davydychev:1995nq}; as well as dimensional shift identities in the \review{parametric representation \cite{Tarasov:1996br,Anastasiou:1999bn,ReFiorentin:2015kri,Jin:2022zcj,Chen:2024xwt}}. In calculations of UV counterterms with the $R^*$-method \cite{Chetyrkin:1982nn,Chetyrkin:1984xa,Smirnov:1986me,Chetyrkin:2017ppe} as implemented with the approach in ref.~\cite{Herzog:2017bjx}, high-rank ($\sim 14$) vacuum tensor integrals were encountered in various 5-loop calculations \cite{Herzog:2017dtz,Herzog:2017ohr,Herzog:2018kwj}.

To achieve this tensor reduction, an efficient method for building projectors for products of metric tensors which makes use of the symmetry properties was proposed in \cite{Herzog:2017ohr,Ruijl:2018poj}. These projectors were extended to the case of transverse metric tensors and, in combination with the van Neerven-Vermaseren basis \cite{vanNeerven:1983vr,Binoth:2002xg,Ellis:2011cr}, a closed-form solution for a general projector performing the reduction of arbitrary tensor integrals, depending also on arbitrary external momenta, was presented by Anastasiou et al.\ in ref.\ \cite{Anastasiou:2023koq} organised in terms of Wick contractions. Further developments on the vacuum projectors, its organisation in terms of an \emph{orbit partition formula}, compact expressions for vacuum projectors with up to 32 Lorentz indices, the extension of the approach to spin indices, and an alternative formulation (the tensor basis is identical to the one first proposed in ref.\ \cite{Anastasiou:2023koq} but is implemented via a transverse decomposition of the integrand) for the case with external momenta were presented in ref.\ \cite{Goode:2024mci}. The purpose of this work is to implement the developments of ref.\ \cite{Goode:2024mci} in the \FORM{} \cite{Ruijl:2017dtg,Tentyukov:2007mu,Vermaseren:2000nd} program \OPITER{} (Orbit-Partition-Improved TEnsor Reduction).

%

The \OPITER{} program is flexible and capable of handling tensors with up to rank $20$.  By factorizing out the $\gamma$-matrices \OPITER{} can also handle spinor indices, and up to 8 independent external momenta. Beyond its application to multi-loop $R^*$ calculations the orbit partition approach should be particularly useful when taking asymptotic expansions in momentum space on a diagram-by-diagram basis \cite{Smirnov:1990rz,Smn94,BnkSmn97,SmnRkmt99,
Gardi:2022khw,Ma:2023hrt,Herzog:2023sgb}. \OPITER{} is available at the following repository: \href{https://bitbucket.org/jaegoode/opiter}{bitbucket.org/jaegoode/opiter}.


The paper is structured as follows: in \cref{sec:orbit_partition_aproach} we review the orbit partition approach and its extension to external momenta via the van Neerven-Vermaseren basis.
In \cref{sec:conventions} we set out the conventions used in the \OPITER{} code and provide examples of how to run it. In \cref{sec:procedures} we discuss the structure of the code, highlighting also the utility of certain procedures. Checks and performance benchmarks are presented in \cref{sec:checks}. Finally, our conclusions are presented in \cref{sec:conclusion}.

\section{Tensor reduction approach}
\label{sec:orbit_partition_aproach}
We wish to reduce $D$-dimensional tensor Feynman integrals of the form
\begin{equation}\label{eq:reduction_fullint}
  \begin{split}
    I^{\mu_1\dots\mu_N} (q_1,\dots,q_E)= \int &  \dd^D p_1\dots \dd^D p_L\,{N^{\mu_1\dots\mu_N}(p_1,\dots ,p_L)}\\ &\times{\mathcal{I}(p_1,\dots,p_L;q_1,\dots,q_E)}\,,
  \end{split}
\end{equation} 
where $N$ is the tensorial part of the numerator and  $\mathcal{I}$ is some scalar function of all the momenta and masses (the dependence on which we suppress) in the problem, $q_1,\,\dots\,,q_E$ are \review{linearly independent} external momenta and $p_1,\,\dots\,,p_L$ are \review{linearly independent} loop momenta. The purpose of \OPITER{} is to reduce integrals of the type in \cref{eq:reduction_fullint} to the form
\begin{equation}\label{eq:reduction_reducedint}
    I^{\mu_1\dots\mu_N}  = I_1 t_1^{\mu_1\dots\mu_N} +I_2 t_2^{\mu_1\dots\mu_N} + \,\cdots\,,
\end{equation} 
where the $I_i$ are some purely scalar integrals and the $t_i$ are some tensor structures independent of the loop momenta. 

\subsection{Transverse tensor reduction}
\label{sec:fullbasis}
The first step in this reduction is to move to the van Neerven-Vermaseren  basis \cite{vanNeerven:1983vr,Binoth:2002xg,Ellis:2011cr} by splitting the $D$-dimensional loop momentum space, $V$, in to a subspace, $V_\parallel$,  spanned by the \review{linearly-independent} external momenta and its transverse complement, $V_\perp$. $V_\parallel$ is a well-defined $E$-dimensional vector space. In doing this we have made the usual assumption of dimensional regularization \cite{tHooft:1972tcz} that $D>E$. Each loop momentum may be decomposed as follows
\begin{equation}
  p_i^\mu = (p_i)_\perp^\mu + (p_i)_\parallel^\mu,
\end{equation}
where $(p_i)_\parallel^\mu$ can only be some linear combination of external momenta which immediately factorises out of the integral. 
We also decompose the $D$-dimensional metric, $g$, into two pieces 
\begin{equation}
    g^{\mu\nu} = g_\perp^{\mu\nu} + g_\parallel^{\mu\nu}.
\end{equation}
$g_\parallel^{\mu\nu}$ may be expressed as: 
\begin{equation}
  g_\parallel^{\mu\nu}= \sum_{i,j} q^\mu_i G^{-1}_{ij}q^\nu_j\,, \quad q_i\cdot q_j = G_{ij}\,,
\end{equation}
\review{where $G_{ij}$ is the usual Gram matrix of the linearly independent external momenta.}
With these definitions we have 
\begin{equation}
  (p_i)_\perp^\mu =(p_i)_\nu\, g_\perp^{\mu\nu}\,, \qquad  (p_i)_\parallel^\mu =(p_i)_\nu\, g_\parallel^{\mu\nu}\,.
\end{equation}

In the following we employ the Schoonschip shorthand, native also to \FORM{}, where contraction with a vector is denoted by placing the vector in the index position, e.g.
\begin{equation}
    T^{\alpha\beta} p_\beta =: T^{\alpha p}.
\end{equation}
It is convenient to introduce dual momenta
\begin{equation}\label{eq:dualdotmom}
  q_i\cdot r_i=\delta_{ij}\,,\quad \text{such that} \quad r_i\cdot r_j= G^{-1}_{ij}\,.
\end{equation}
The dual momenta can be expressed as \cite{Ellis:2011cr}
\begin{equation}\label{eq:dualdeff}
  r_i^\mu = \frac{\delta_\parallel{}^{q_1 \dots q_{i-1}\,\mu\,q_{i+1}\dots q_E}_{q_1\quad\dots\quad q_E}}{\Delta(q_1\dots q_E)}\,, \qquad \Delta(q_1\dots q_E) = \delta^{q_1 \dots  q_E}_{q_1\dots q_E}\,,
\end{equation}
with the generalised Kronecker delta defined by
\begin{equation}
  \delta_{\nu_1 \dots \nu_E}^{\mu_1 \dots \mu_E}=\left|\begin{array}{ccc}
    \delta_{\nu_1}^{\mu_1} & \cdots & \delta_{\nu_E}^{\mu_1} \\
  \vdots & \ddots & \vdots \\
  \delta_{\nu_1}^{\mu_E} & \cdots & \delta_{\nu_E}^{\mu_E}
  \end{array}\right| \quad \text{or} \quad \delta_{\nu_1 \dots \nu_E}^{\mu_1 \dots \mu_E}=p!\delta_{\left[\nu_1\right.}^{\mu_1} \dots \delta_{\left.\nu_E\right]}^{\mu_E}\,,
  \end{equation}
and its restriction to $V_{\parallel}$ \footnote{Strictly speaking we should say $V_{\parallel}\otimes V_{\parallel}\otimes V_{\parallel}\dots$.}, denoted by $\delta_\parallel{}$, given by replacing $\delta^{\mu}_\nu$s with $\delta_{\parallel}{}^{\mu}_\nu$s. 
We may convert between the external and dual  bases using 
\begin{equation}\label{eq:momentumtransform}
  q_i^\mu = \sum_{j=1}^E G_{ij}\, r^\mu_j\,,\qquad \text{and} \qquad r_i^\mu = \sum_{j=1}^E G^{-1}_{ij}\, q^\mu_j\,.
\end{equation}
Note also that
\begin{equation}
  \delta_\parallel{}^{\mu_1\dots \mu_E}_{\nu_1\dots \nu_E} = \epsilon_\parallel{}^{\mu_1\dots \mu_E}\epsilon_\parallel{}_{\nu_1\dots \nu_E}\,,
\end{equation}
where $\epsilon_\parallel{}^{\mu_1\dots \mu_E}$ \review{is the Levi-Civita symbol} defined in the $E$-dimensional subspace $V_\parallel$. We define $\epsilon_\parallel{}^{\mu_1\dots \mu_E}=0$ for values of the Lorentz indices $\mu_i$ not in $V_\parallel$.
Equipped with this identity we may express the dual momenta
\begin{equation}
\begin{split}
  r_i^\mu  &= 
  \frac{\epsilon_\parallel{}^{q_1 \dots q_{i-1}\,\mu\,q_{i+1}\dots q_E}}{\epsilon_\parallel{}^{q_1\dots q_E}}\,.
  \end{split}
\end{equation}
Now we may use this to express the elements of the inverse Gram matrix  
\begin{equation} \label{eq:inversGramElement}
  \begin{split}
    G^{-1}_{ij}  &= \frac{\epsilon_\parallel{}^{q_1 \dots q_{i-1}\,\mu\,q_{i+1}\dots q_E}\epsilon_\parallel{}_{q_1 \dots q_{j-1}\,\mu\,q_{j+1}\dots q_E}}{\epsilon_\parallel{}^{q_1\dots q_E}\epsilon_\parallel{}_{q_1\dots q_E}}\\
    &=\frac{\delta{}^{q_1 \dots q_{i-1}\,q_{i+1}\dots q_E}_{q_1 \dots q_{j-1}\,q_{j+1}\dots q_E}}{\Delta(q_1\dots q_E)}\,(-1)^{i+j}\,,
  \end{split}
\end{equation}
where in the last line we use the antisymmetry of the generalised Kronecker delta and the following identity  
\begin{equation}
  \delta_\parallel{}^{\mu_1 \dots \mu_s\,\mu_{s+1}\dots \mu_p}_{\nu_1 \dots \nu_s\,\mu_{s+1}\dots \mu_p} = \frac{(n-s)!}{(n-p)!}\delta_\parallel{}^{\mu_1 \dots \mu_s}_{\nu_1 \dots \nu_s}\,.
\end{equation}
We are also free to drop the parallel requirement on the Kronecker delta as everything is contracted with external momenta.
We are now free to apply the decomposition 
\begin{equation}\label{eq:loopdecomp}
    p_i^\mu = (p_i)_\perp^\mu + \sum_{j=1}^{E} p_{i}\cdot q_{j}\, r^\mu_j \,,
\end{equation}
to the integral in \cref{eq:reduction_fullint}.

After expanding, this fully factorises all the external momenta dependent parts of the tensor structures and so only vacuum tensors living in the transverse space remain in the integral. 
Let us now focus on a single term in the expansion with all the dual momenta  $r_i^\mu$ stripped. We can express such a purely transverse integral on a basis of products of metric tensors, i.e. structures of the form
\begin{equation}
\label{eq:Tiperp}
    t_{\perp}^{\mu_1 \dots \mu_N} (\sigma) = g_\perp^{\mu_{\sigma(1)}\mu_{\sigma(2)}} \dots g_\perp^{\mu_{\sigma(N-1)}\mu_{\sigma(N)}}\,,
\end{equation}
where the $\sigma$ is some permutation in the set $S^N_2$ which generates all possible distinct $t_\perp(\sigma)$. For each element $t_\perp(\sigma)$ there then exists a projector
$P_\perp(\sigma)$, such that
\begin{equation}
    P_\perp(\sigma)\cdot t_\perp(\sigma') = \delta_{\sigma\sigma'},
\end{equation}
where the central dot represents a full contraction of all indices $A\cdot B = A^{\mu_1\dots \mu_N}B_{\mu_1\dots \mu_N}$. The $P_\perp(\sigma)$ were computed via the orbit partition approach up to rank 32 in ref.~\cite{Goode:2024mci}. Concretely, given some integral (we ignore the denominator since it doesn't participate in the reduction),
\begin{equation}
   I^{\mu_1\dots\mu_N}= \int \dd^D p_1 \dots \dd^D p_L\, N^{\mu_1\dots\mu_N}(p_1,\dots ,p_l),
\end{equation}
we can reduce it onto the $t_\perp(\sigma)$ basis by applying the projectors. This leaves us with:
\begin{equation}\label{eq:fullbasis:tensorbasis}
    \begin{split}
        I^{\mu_1\dots\mu_N} &=\sum_{\sigma\in S^N_2}\, t_\perp^{\mu_1\dots\mu_n}(\sigma)  \int \dd^D p_1 \dots \dd^D p_L\ P_\perp(\sigma) \cdot N\,,\\
        &= \sum_{\sigma\in S^N_2}\, t_\perp^{\mu_1\dots\mu_n}(\sigma) I(\sigma)\,.
    \end{split}
\end{equation}
In the case that the integral contains spinors we split the slashed momenta up, via (schematically)
\begin{equation}
\int \dd^D p \, \dots  \slashed{p}\dots \,{=}\, \gamma_\mu \int \dd^D p \, \dots p^\mu\dots  \,,
\end{equation}
and the tensor reduction is then performed as above on the pure Lorentz structure only, making use of
the basis of antisymmetrised $\Gamma$-matrices to reduce the number of integrals; see \cref{sec:gamma2Gamma}.

\subsection{Integrand symmetries} \label{sec:intSym}

A problem faced at high tensor rank $N$ is that the number of terms in the basis in \cref{eq:fullbasis:tensorbasis} grows factorially. A way to tame this growth is to take advantage of the symmetries, present in the integrand under exchanges of Lorentz indices, which cause many of the scalar integrals, $I(\sigma)$, to be equal. In the following we present a method to build a basis of tensors which manifests integrand symmetries.

Let $\sum_{i=1}^L N_i=N$. Introduce $N$ indices $\mu_{i,j}$ such that $i\in \{1,\dots,L\}$ and $j\in \{1,\dots, N_i\}$ for a given $i$.
A general transverse integral is then given by
\begin{equation}
\label{eq:twolooptensorN}
I_\perp^{\vec{\mu}} =\int  \left( \prod_{i=1}^L d^Dp_i \prod_{j=1}^{N_i} p_\perp^{\mu_{i,j}} \right) \mathcal{I}(p_1,\dots,p_L; q_1,\dots,q_E)\,,
\end{equation}
where we introduced the shorthand
$$
\vec{\mu}=\mu_{1,1}\dots \mu_{1,N_1}\dots\mu_{L,1}\dots \mu_{L,N_L}.
$$
This integrand has a symmetry which is given by the product group $H=S_{N_1}\times S_{N_2}\times \cdots \times S_{N_L}$. In certain cases there could be more symmetry due to re-parameterisation symmetries of the loop momenta $p_i$ of the scalar integrand $\mathcal{I}$, or even integral $I$. We do not take such additional symmetries into account in the following discussion.
Because of the symmetry group $H$ many coefficients of the different $t_{\perp}$-structures, \cref{eq:Tiperp}, appearing in the tensor-reduced form of \cref{eq:twolooptensorN} will be identical. The efficiency of the tensor-reduction algorithm can profit enormously by taking this symmetry into account in the construction process. In particular we need to partition the different possible $t_{\perp}$-structures into sums invariant under $H$. It is now convenient to introduce the  totally symmetric transverse tensor,
\begin{equation}
d_{\perp}^{\mu_1\dots\mu_N}= \sum_{\sigma\in S_{2}^N} t_{\perp}^{\mu_1 \dots \mu_N}(\sigma)\,.
\end{equation}
The $H$-invariant sums can be generated by contracting
$d_{\perp}^{\vec{\mu}}$
with the tensorial part of the integrand of \cref{eq:twolooptensorN}. This results in the  equation
\begin{equation}
\label{eq:dperpcontracted}
d_\perp^{\,\overbrace{p_1\dots p_1}^{N_1}  \overbrace{p_2\dots p_2}^{N_2}
\dots \overbrace{p_L\dots p_L}^{N_L}}
=\sum_\alpha c(\alpha) m(\alpha)\,,
\end{equation}
where $c(\alpha)$ are positive integers counting the appearances of different monomials,
\begin{equation}
m(\alpha)=\prod_{i\le j}  (p_i.p_j)^{\alpha_{ij}}\,,
\end{equation}
with $\alpha$ a matrix. The elements are defined such that $\alpha_{ij}=\alpha_{ji}$.
Furthermore, $\alpha$ must satisfy the following constraints:
\begin{equation}
  \sum_{i=1}^L\sum_{i\le j} \alpha_{ij}=N/2\,, \qquad \sum_{j=1}^L \alpha_{ij} + \alpha_{ii} = N_i\,.
\end{equation}
The combinatorial factor $c(\alpha)$ is given by
\begin{equation}
    \begin{split}
    c(\alpha) =& \prod_{i=1}^L \prod_{j=i+1}^L \frac{\Big(N_{i}\Big)_{2\alpha_{ii}}\left(n_{i,j}\right)_{\alpha_{ij}}\left(n_{j,i}\right)_{\alpha_{ij}}}{\,2^{\alpha_{ii}}{\alpha_{ii}}!\,\alpha_{ij}!}\,,\quad n_{i,j}=\!\!\! \sum_{k=i,\, k\neq j}^L\!\!\alpha_{kj}\,,
    \end{split}
\end{equation}
where $(x)_n = \frac{x!}{(x-n)!}$ is the Pochhammer symbol for the falling factorial. A derivation of this factor is presented in \ref{sec:caproof}.
We see that different contractions of the tensorial part of the integrand are either identical or not. Now, because the projectors $P_{\perp}(\sigma)$ have the same symmetry properties as the
``contractions'' $t_{\perp}(\sigma)$, we can deduce that if two monomials generated from two $t_{\perp}$ are the same, then the corresponding two projectors acting on the integrand will also yield the same result. Therefore every monomial $m(\alpha)$ corresponds to a different $H$-invariant sum labelled by a particular matrix $\alpha$ (a more complete proof of this is presented in \ref{sec:Hinvarc}).

We thus arrive at the equation
\begin{align}
I_\perp^{\vec{\mu}}
 &=\sum_\alpha I(\alpha) T_\perp^{\vec{\mu}}(\alpha)\,,\qquad I(\alpha) =P_\perp(\alpha) \cdot I_\perp\,,
\end{align}
where $T_\perp(\alpha)$ are the distinct $H$-invariant tensors, $I(\alpha)$ are their corresponding scalar integrals and $P_\perp(\alpha)$ is a projector for any $t_{\perp}(\sigma)$ appearing in the $H$-invariant sum $T_\perp(\alpha)$. An explicit expression defining $T_\perp(\alpha)$ is given by
\begin{equation}
\label{eq:Tdiffrep}
T_\perp^{\vec{\mu}}(\alpha)= \frac{c(\alpha)}{N_1!\cdots N_L!}
\left(\prod_{i=1}^L
\frac{\partial^{N_i} }{\partial p_i^{\mu_{i,1}}\cdots \partial p_i^{\mu_{i,N_i}}}\right)\, m(\alpha)\,.
\end{equation}
A proof for this expression is presented in \ref{sec:formT}. In practice we generate the $T_\perp(\alpha)$ by contracting the loop momenta with $d_\perp$. This is very efficiently done in \FORM{} though the \lstinline|dd_| function. We may then extract an element $t_\perp(\alpha)$ to generate $T_\perp(\alpha)$ by replacing each momentum with an index that momentum carried in the integrand and the dot products with metrics.
The exact choice of $t_\perp(\alpha)$ is not unique but ultimately unimportant.
To generate the full $T_\perp(\alpha)$ we express the index symmetry of the integrand in terms of symmetrisers acting on this generating element.
A generic one has the form
\begin{equation}\label{eq:symmetriserdef}
  \mathcal{S}^{\mu_1\dots\mu_N}_{\nu_1\dots\nu_N} =\delta^{\mu_1}_{(\nu_{\sigma(1)}}\dots \delta^{\mu_{N}}_{\nu_{\sigma(N)})} =\frac{1}{N!}\sum_{\sigma\in S_{N}}\delta^{\mu_1}_{\nu_{\sigma(1)}}\dots \delta^{\mu_{N}}_{\nu_{\sigma(N)}}\,,
\end{equation}
so an invariant tensor of the integral in \cref{eq:twolooptensorN} would be
\begin{equation} \label{eq:Talphasym}
  T_\perp^{\vec{\mu}}(\alpha)=
  c(\alpha)\, \left(\prod_{i=1}^L \mathcal{S}^{\mu_{i,1}\dots\mu_{i,N_i}}_{\nu_{i,1}\dots\nu_{i,N_i}}\right) \, t_\perp^{\nu_{1,1}\dots \nu_{1,N_1}\dots\nu_{L,1}\dots \nu_{L,N_L}}(\alpha)\,.
\end{equation}
The evaluation of such symmetrisers is described in \cref{sec:symmetrizer}.

\section{Conventions and running the code}
\label{sec:conventions}
\subsection{Conventions and setup}\label{sec:conventions-conventions-and-setup}
\OPITER{} is presented as a collection of \FORM{} procedures in a manner that is intended to maximise ease of integration into existing projects. 
Before using \OPITER{} you must load the procedures into \FORM{}'s search path.
To do this, include the following lines at the top of your \FORM{} program:
\begin{lstlisting}
#: IncDir opiter    
#include opiter.frm
\end{lstlisting}
Alternatively, one can add the line \lstinline|IncDir opiter| to the setup file (\lstinline|form.set| by default).
\FORM{} has many such setup options to control multithreading, memory allocation, and so on. 
Please see the \FORM{} documentation for a complete description.

\OPITER{} exposes several options to the user. 
These are controlled by the preprocessor variables defined in \lstinline|opiter/opitersettings.dat|. 
The first of these is \lstinline|tensormode|. If \lstinline|tensormode=2| the integrand symmetry simplifications described in  \cref{sec:intSym} are enabled. Such simplifications are instead ignored if \lstinline|tensormode=1|. The second preprocessor variable is \lstinline|tensorbasis|. For \lstinline|tensorbasis=1| the results of the reduction are expressed in terms of the dual-transverse basis (in terms of $g_\perp$ and $r_i^\mu$). Alternatively for \lstinline|tensorbasis=2| the results are presented in the standard basis ($g$ and external momenta).

To illustrate the effect of these settings we have included the file \lstinline|example.frm| which is intended as a minimal implementation and usage of the \OPITER{} procedure.
More information on the example file is given in \cref{sec:example}.

Projectors with large numbers of Lorentz indices will take some time to load in so we provide the option to leave out projectors that are not needed. \OPITER{} comes pre-set with recommended values of \lstinline|maxE|, the maximum number of external momenta $E$, and \lstinline|maxN|, the maximum number of tensor rank $N$, that determine which projector files will be loaded in; these numbers can be increased up to 8 and 20 respectively.

\OPITER{} acts on active \FORM{} expressions in the current program. 
However, there is some extra metadata you need to attach to the input expression in order to tell \OPITER{} which momenta are external and which are loop.
This is achieved through the \lstinline|ext| and \lstinline|loop| helper functions which contain a list of their respective momenta.
Thus a sample input would look like the following:
\begin{lstlisting}
L F = ext(q1,q2)*loop(p1,p2)
    *p1(mu1)*p1(mu2)*p2(mu3)*p2(mu4);
\end{lstlisting}
\review{Note that only linearly independent momenta should enter the \lstinline|ext|-function, since otherwise the corresponding Gram determinant would vanish and its inverse be singular, meaning that the tensor-reduced expression could not be evaluated numerically. Note also that when one wishes to set $D=4$ in the tensor reduced expressions (assuming integrals have been performed and poles cancelled) that no more than 4 independent momenta should enter the \lstinline|ext|function.}

The following functions/symbols have special meaning and can be used in your input file:

\noindent \codeparaheading{gam(line,indices)}: This function defines a string of gamma matrices using the same convention as \FORM{}'s own \lstinline|g_|. Alternatively one can use \lstinline|Gsigma| which denotes a totally antisymmetric product of gamma matrices.

\noindent \codeparaheading{rat(x,y)}: During the runtime of \OPITER{} we assign \lstinline|rat| to be the \lstinline|PolyRatFun| (polynomial rational function), meaning that \FORM{} will combine and simplify neighbouring \lstinline|rat|s.
More information on the behaviour of \lstinline|PolyRatFun| can be found in the \FORM{} documentation.

\noindent \codeparaheading{proj(...)}: It may be that your integrand contains a number of external momenta present only in the numerator. In this case these momenta can be factored out of the integral, in the form of a \emph{projector} to be acted on the tensor integral after tensor reduction. This is a typical situation faced when Taylor expanding integrals in external momenta. By feeding this product of external momenta into a \lstinline|proj| function \OPITER{} can exploit the symmetry to make the
\lstinline|symmetrise| procedure more efficient. We discuss \lstinline|proj| further in \cref{sec:symmetrizer}.

\noindent \codeparaheading{deno(x)}: Since \OPITER{} is a tensor reduction routine it does not care about anything in the denominator of the Feynman integral being reduced.
\lstinline|deno(x)| is simply a shorthand for $1/x$ that also protects its contents from any manipulation by the \OPITER{} procedure.
It may also appear in your output.

\subsection{Running the code}
Once the input is defined, the \OPITER{} procedure is invoked through the following command:
\begin{lstlisting}
#call opiter
\end{lstlisting}
and tensor reduction will be performed on any currently active \FORM{} expressions.
Depending on the settings in \lstinline|opiter/opitersettings.dat|, there may be unexpanded symmetrisers and dual momenta in your output.
Procedures are provided so that the user can expand these quantities later.
To expand out the symmetrisers:
\begin{lstlisting}
#call symmetrise
\end{lstlisting}
To move back to the non-dual-transverse basis:
\begin{lstlisting}
#call leavedualtransverse
\end{lstlisting}

\subsection{Understanding the output}
\label{sec:understanding-output}
The result of the \lstinline|opiter| procedure is a fully tensor-reduced expression consisting of scalar integrals that can be used in the next step of your computation.
In addition to the functions described in \cref{sec:conventions-conventions-and-setup} there are some further functions that can appear in the output of the \OPITER{} procedure.

\noindent \codeparaheading{dual(qi,index)}: This denotes the dual momentum corresponding to the external momentum \lstinline|qi|, as defined in \cref{eq:dualdotmom}.
In that section they were denoted by $r_i$.

\noindent \codeparaheading{sym(ind1(...)*ind2(...))}: Denotes a symmetriser $\mathcal{S}$ as defined in \cref{eq:symmetriserdef}. The two index sets in \lstinline|ind1| and \lstinline|ind2| correspond to the upper and lower set of indices.

\noindent \codeparaheading{dts, ddts}: Both refer to the transverse metric $g_\perp$ discussed throughout \cref{sec:orbit_partition_aproach}.

\subsection{Reserved symbols}
\FORM{}'s lack of encapsulation makes it impossible to hide the symbols we use in our procedures from the end user. We have endeavoured to use symbols that are not likely to clash with those in the user's own programs. However, we reserve certain symbols for internal use in \OPITER{}; please see the \lstinline|README.md| file in the \OPITER{} repository for a complete and up-to-date list.
%
%
%

\label{sec:example}
\subsection{The example file}
The file \lstinline|example.frm| in the \OPITER{} repository is intended to show a minimal invocation of \OPITER{} along with the output of some simple tensor reduction examples.
These are as follows:
\begin{enumerate}
    \item $p_1^{\mu_1} p_2^{\mu_2} p_2^{\mu_3} p_2^{\mu_4}$
    \item $p_1^{\mu_1} p_2^{\mu_2}$
    \item $p_1^{\mu_1} \slashed{p}_2 \gamma^{\mu_2}$
    \item $p_1^{\mu_1}\dots p_1^{\mu_5} p_2^{\mu_6} \dots p_2^{\mu_{10}}$.
\end{enumerate}

First of all, we must include the following to bring all of \OPITER{}'s procedures into the \FORM{} search path:
\begin{lstlisting}
#: IncDir opiter/
#include- opiter.frm
\end{lstlisting}

Following this we set up four local expressions to correspond to the four examples:
\begin{lstlisting}
L F1a = ext()*loop(p1,p2)
         *p1(mu1)*p2(mu2)*p2(mu3)*p2(mu4);
L F1b = F1a;

L F2 = ext(q1,q2)*loop(p1,p2)
         *p1(mu1)*p2(mu2);

L F3 = ext(q1)*loop(p1,p2)
         *p1(mu1)*gam(1,p2,mu2);

L F4 = proj(<q1(mu1)>*...*<q1(mu10)>)
         *ext()*loop(p1,p2)
         *<p1(mu1)>*...*<p1(mu5)>
         *<p2(mu6)>*...*<p2(mu10)>;
\end{lstlisting}
In each, the external momenta are listed inside the \lstinline|ext| function, and the loop momenta in the \lstinline|loop| function. 
The momenta and indices in the expressions have been declared using an \lstinline|AutoDeclare| statement. 
Furthermore, the third example shows the use of \lstinline|gam| to declare gamma matrices with contracted (i.e.\ slashed) momenta.
The fourth example involves the \lstinline|proj| function which was introduced in \cref{sec:conventions-conventions-and-setup}.
For compactness we also use \FORM{}'s triple dot notation which simply fills in the missing values.
For instance \lstinline|<q1(mu1)>*...*<q1(mu10)>| fills in \lstinline|q1(mu1)*q1(mu2)*...|, up to \lstinline|q10(mu10)|.

Next, we call the \lstinline|opiter| procedure:
\begin{lstlisting}
#call opiter
\end{lstlisting}
The procedure works on all active expressions, so \lstinline|F1|, \dots, \lstinline|F4| will all be tensor reduced.

According to the default settings in \lstinline|opitersettings.dat|, the symmetrisers that result from the tensor reduction are left unexpanded for the sake of compactness.
Calling the \lstinline|symmetrise| procedure expands these symmetrisers out fully:
\begin{lstlisting}
#call symmetrise
\end{lstlisting}
To demonstrate the difference, we \lstinline|hide F1a| before doing so, which leaves the symmetriser in \lstinline|F1a| unexpanded.

After running \lstinline|example.frm| we can inspect the output. 
For the first example we obtain (the output format has been altered for readability):
\begin{lstlisting}[lastline=7]
F1a =
   + ddts(MMu2,mu1)*ddts(MMu3,MMu4)
   * sym(
       ind1(mu2,mu3,mu4)
      *ind2(MMu2,MMu3,MMu4)
   )
   *ext*loop(p1,p2)*p1.p2*p2.p2*rat(3,D^2+2*D)
\end{lstlisting}
We see that the expression has been tensor reduced, with the Lorentz indices now living inside the \lstinline|ddts| functions (the transverse metric tensors) and \lstinline|sym| functions. The unexpanded symmetriser in \lstinline|F1a| follows the syntax described in \cref{sec:understanding-output}.
In \lstinline|F1b| the symmetriser is allowed to be expanded, resulting in three terms,
\begin{lstlisting}[firstline=9, lastline=17]
F1b =
  + ddts(mu1,mu2)*ddts(mu3,mu4)*ext*loop(p1,p2) 
     * p1.p2*p2.p2*rat(1,D^2 + 2*D)

  + ddts(mu1,mu3)*ddts(mu2,mu4)*ext*loop(p1,p2) 
     * p1.p2*p2.p2*rat(1,D^2 + 2*D)

  + ddts(mu1,mu4)*ddts(mu2,mu3)*ext*loop(p1,p2) 
     * p1.p2*p2.p2*rat(1,D^2 + 2*D)
\end{lstlisting}
which are evidently symmetric on the relevant indices.

Example 2 shows a simpler example but in the presence of some external momenta.
The result of the reduction is:
\begin{lstlisting}[firstline=19,lastline=35]
F2 =
  + ddts(mu1,mu2)*ext(q1,q2)*loop(p1,p2)
     *deno(q1.q1*q2.q2 - q1.q2^2) * (
     + p1.q1*p2.q1*q2.q2*rat(-1,D - 2)
     + p1.q1*p2.q2*q1.q2*rat(1,D - 2)
     + p1.q2*p2.q1*q1.q2*rat(1,D - 2)
     + p1.q2*p2.q2*q1.q1*rat(-1,D - 2)
     )

  + ddts(mu1,mu2)*ext(q1,q2)*loop(p1,p2) 
     * p1.p2*rat(1,D - 2)

  + ext(q1,q2)*loop(p1,p2)
     *dual(q1,mu1)*dual(q1,mu2) 
     * p1.q1*p2.q1*rat(1,1)

*  + other terms ...
\end{lstlisting}
Notable here is the appearance of \lstinline|dual|s in the output. 
These represent the dual momenta which were denoted $r_i$ throughout \cref{sec:fullbasis}.
The \lstinline|leavedualtransverse| procedure is provided to convert these back into ordinary momenta or, alternatively, one can change the \lstinline|tensormode| in \lstinline|opitersettings.dat| to do this automatically.

\lstinline|F3| uses \lstinline|gam| in the input to represent gamma matrices:
\begin{lstlisting}[firstline=49, lastline=58]
F3 =
  + ddts(muro1,mu1)*Gsigma(1,muro1,mu2)
     *ext(q1)*loop(p1,p2) * (
     + p1.p2*rat(1,D - 1)
     + p1.q1*p2.q1*q1.q1^-1*rat(-1,D - 1)
     )

  + Gsigma(1,muro1,mu2)*ext(q1)*loop(p1,p2)
     *dual(q1,muro1)*dual(q1,mu1) 
     *p1.q1*p2.q1*rat(1,1);
\end{lstlisting}
The symbol appearing in the output is \lstinline|Gsigma|, the totally antisymmetric product of gamma matrices.

\lstinline|F4| is meant to show the functionality of the \lstinline|proj| function, which is explained in \cref{sec:conventions-conventions-and-setup}. 
The output of the procedure is of the following form:
\begin{lstlisting}[firstline=61, lastline=66]
F4 =
  + ddts(q1,q1)^5*ext*loop(p1,p2) * (
     + p1.p1*p1.p2^3*p2.p2*rat(...)
     + p1.p1^2*p1.p2*p2.p2^2*rat(...)
     + p1.p2^5*rat(...)
     );
\end{lstlisting}
The arguments of the \lstinline|rat| functions are polynomials in $D$ of fairly high degree and have been omitted for compactness. The arguments of the \lstinline|proj| function had the effect of collapsing all rank 10 tensor structures into the single term \lstinline|ddts(q1,q1)^5| on the fly.

\section{Program structure and procedures}
\label{sec:procedures}
In this section we discuss the general structure of the code and highlight several useful stand-alone procedures.
The program operates in the following steps:
\begin{enumerate}
  \item $\gamma$-matrices are decomposed into the antisymmetric basis (see \cref{sec:gamma2Gamma}).
  \item  Loop momenta are decomposed into the van Neerven-Vermaseren basis.
  \item Tensor reduction is performed on the remaining $(p_i)_\perp^\mu$ through the methods outlined in \cref{sec:fullbasis} and \cref{sec:intSym}.
  \item The contracted projectors are evaluated.
  \item The loop momenta and $g_\perp$ are transformed back if requested.
\end{enumerate}

We now introduce some procedures. Some of these may be useful in their own right also beyond \OPITER{}. Others include novel \FORM{} programming methods,
which could be useful in other situations and thus deserve highlighting.

\subsection{Antisymmetric basis for $\gamma$-matrices} \label{sec:gamma2Gamma}
The first basis transformation we perform is transforming into the antisymmetric basis of gamma matrices. The antisymmetric gamma matrices are defined as
\begin{equation}\label{eq:antisymmetric-gamma}
    \Gamma^{\mu_1\dots\mu_p}=\gamma^{[\mu_1}\dots\gamma^{\mu_p]}= \frac{1}{p!}\,\delta^{\mu_1\dots\mu_p}_{\nu_1\dots\nu_p}\gamma^{\nu_1}\dots\gamma^{\nu_p}\,,
\end{equation}
form a basis free of Clifford algebra relations, and satisfy a useful orthogonality property \cite{Kennedy:1981kp}. There is an efficient way to convert a product of gamma matrices into the basis of $\Gamma$s, which we have implemented in the procedure \verb|gamma2Gamma|. It makes use of the identity
\begin{equation}
    \label{eq:gammatoGamma}
    \gamma^{\mu_1}\dots\gamma^{\mu_n}=\sum_{k=0}^{n}\sum_{\pi\in \Sigma_n^k}\sgn(\pi)\Gamma^{\mu_{\pi{(1)}}\dots\mu_{\pi(k)}}\tr(\gamma^{\mu_{\pi(k+1)}}\dots\gamma^{\mu_{\pi(n)}})\,,
\end{equation}
where the sum over $\Sigma_{n}^k$ shuffles the first $k$ indices with the remaining $n-k$ indices over the two tensors \cite{Goode:2024mci}. This is simple to implement in \FORM. The indices of the product of gamma matrices on the left are split between the $\Gamma$ and the trace and are then permuted with the appropriate sign using \FORM's \verb|distrib_| function. The trace is then done with \FORM's built-in implementation of gamma matrices.
The simple example
\begin{lstlisting}
L F = gam(1,mu1,mu2,mu3,mu4);

#call gamma2Gamma
\end{lstlisting}
would output 
\begin{lstlisting}
F =
  + Gsigma(1,mu1,mu2,mu3,mu4)*rat(1,1)
  + Gsigma(1,mu1,mu4)*d_(mu2,mu3)*rat(1,1)
  + Gsigma(1,mu2,mu4)*d_(mu1,mu3)*rat(-1,1)
  + Gsigma(1,mu3,mu4)*d_(mu1,mu2)*rat(1,1)
  ;
\end{lstlisting}
where the \lstinline|Gsigma| are the $\Gamma$ defined in \cref{eq:antisymmetric-gamma}. The \lstinline|1| labels spin indices and informs the program if the \lstinline|gam| or \lstinline|Gsigma| belong to the same fermion line. This mirrors the use of \FORM's \lstinline|g_|; see the reference manual for more details.

\review{Note that we work in conventional dimensional regularization \cite{tHooft:1972tcz} where $\gamma_5$ is not rigorously defined. To maximise flexibility the treatment of $\gamma_5$ is left to the user. For the purpose of tensor reduction, our approach is always applicable since the $\gamma_5$ can always be factored out of the integral before calling \OPITER, if necessary by splitting up strings of $\gamma$s. Note that \FORM's inbuilt  $\gamma_5$ (\lstinline|g5_(1)|), whose use is restricted to $D=4$,  is not supported in \OPITER.}

\subsection{ van Neerven-Vermaseren basis}
The transformation into the van Neerven-Vermaseren basis is performed as part of the tensor reduction stage. In \verb|tenred| and \verb|tenredisym| loop momenta are decomposed by applying \cref{eq:loopdecomp}.

To transform back from the transverse metric we apply
\begin{align}
        g_\perp^{\mu\nu} &=  g^{\mu\nu} -\sum_{i,j=1}^E  q_i{}^\mu\, (G^{-1})_{ij}\,  q_j{}^\nu\,,\\
        {p_i}_\perp\cdot{p_j}_\perp &= {p_i}\cdot{p_j} -\sum_{l,m=1}^E  p_i\cdot q_l\, (G^{-1})_{lm}\,  q_m\cdot p_j\,.
\end{align} 
This is done by calling \verb|expanddt|. The transformation of the dual momenta $r_i$ is done by calling \verb|dual2ext| which applies \cref{eq:momentumtransform}. The elements $G^{-1}_{ij}$ are substituted by calling \verb|subinvgram| and then given in terms of the $\Delta$ defined in \cref{eq:dualdeff} which is substituted with \verb|subinvgramdet|. \review{Example input and output for these individual procedures can be found in \ref{sec:procexample}.
For the user's convenience all the of these steps are collected into the single procedure \lstinline|leavedualtransverse|}. 
A minimal example is 
\begin{lstlisting}
L F = dt(mu1,mu2)*dual(q1,mu3)*
      loop(p1,p2)*ext(q1,q2);

#call leavedualtransverse

Bracket ext,loop,deno;
\end{lstlisting}
which has output 
\begin{lstlisting}
F =
+ext(q1,q2)*loop(p1,p2)*
 deno(q1.q1*q2.q2-q1.q2^2)^2*(
  +q1(mu1)*q1(mu2)*q1(mu3)*q2.q2^2*rat(-1,1)
  +q1(mu1)*q1(mu2)*q2(mu3)*q1.q2*q2.q2*rat(1,1)
  +q1(mu1)*q1(mu3)*q2(mu2)*q1.q2*q2.q2*rat(1,1)
  +q1(mu1)*q2(mu2)*q2(mu3)*q1.q2^2*rat(-1,1)
  +q1(mu2)*q1(mu3)*q2(mu1)*q1.q2*q2.q2*rat(1,1)
  +q1(mu2)*q2(mu1)*q2(mu3)*q1.q2^2*rat(-1,1)
  +q1(mu3)*q2(mu1)*q2(mu2)*q1.q1*q2.q2*rat(-1,1)
  +q2(mu1)*q2(mu2)*q2(mu3)*q1.q1*q1.q2*rat(1,1)
  )
+ext(q1,q2)*loop(p1,p2)*
 deno(q1.q1*q2.q2 - q1.q2^2) * (
   +d_(mu1,mu2)*q1(mu3)*q2.q2*rat(1,1)
   +d_(mu1,mu2)*q2(mu3)*q1.q2*rat(-1,1)
  );
\end{lstlisting}
\review{Currently \OPITER{} is capable of handling up to 8 independent momenta. The only restriction on this is the availability of the inverse-Gram matrix elements $G^{-1}_{ij}$. If required this could be extended upon request to the authors.}

\subsection{The symmetriser}\label{sec:symmetrizer}
If using the integrand symmetry mode, \OPITER{} will output the tensors in terms of a generating term contracted with various symmetrisers (see \cref{eq:Talphasym}). 
These symmetrisers are represented in \OPITER{} by \lstinline|sym(ind1(mu1,mu2,...)*ind2(nu1,nu2,...))| acting on an expression with indices \lstinline|nu1,nu2,...| to be symmetrised. The procedure \lstinline|symmetrise| can be called to efficiently expand them. The procedure works by decomposing longer symmetrisers into smaller ones. This is done iteratively by shuffling (or cycling) the first index  \lstinline|nu1| with the remaining indices in all possible ways, and then repeating the procedure for \lstinline|nu2|, and so on. After each shuffle, the expression is sorted to allow for simplifications to occur. A minimal example is
\begin{lstlisting}
L F = sym(ind1(mu1,mu2)*ind2(MMu1,MMu2))*
      sym(ind1(mu3,mu4)*ind2(MMu3,MMu4))*
      q1(MMu1)*q2(MMu2)*q3(MMu3)*q4(MMu4);

#call symmetrise  
\end{lstlisting}
which has output 
\begin{lstlisting}
F =
     + q1(mu1)*q2(mu2) * (
        + q3(mu3)*q4(mu4)*rat(1,4)
        + q3(mu4)*q4(mu3)*rat(1,4)
        )
     + q1(mu2)*q2(mu1) * (
        + q3(mu3)*q4(mu4)*rat(1,4)
        + q3(mu4)*q4(mu3)*rat(1,4)
        );
\end{lstlisting}
where it is clear that the result is symmetric in exchanging \lstinline|mu1| and \lstinline|mu2| as well as \lstinline|mu3| and \lstinline|mu4|. Additionally, if the integral is contracted with something symmetric, \lstinline|symmetrise| may use this information to take advantage of simplifications already at the generation stage. These projectors are represented in the \FORM{} code as \lstinline|proj(...)|; an example of this is
\begin{lstlisting}
L F = proj(Q(mu1)*Q(mu2)*Q(mu3)*Q(mu4))*
      sym(ind1(mu1,mu2)*ind2(MMu1,MMu2))*
      sym(ind1(mu3,mu4)*ind2(MMu3,MMu4))*
      q1(MMu1)*q2(MMu2)*q3(MMu3)*q4(MMu4);

#call symmetrise  
\end{lstlisting} 
which directly outputs
\begin{lstlisting}
F =
   + Q.q1*Q.q2*Q.q3*Q.q4*rat(1,1)
;
\end{lstlisting}
without generating the intermediary terms.

\subsection{Projector Contractions}
A crucial component of \OPITER{} is the way in which it performs integrand contractions of the projectors $P_\perp^{\mu_1\dots\mu_n}$. This is done via the procedure
\lstinline|PrtCanonicalize|. This procedure first brings the contracted projectors, e.g. $P_\perp^{p_4p_2p_1p_1}$, into a canonical (or at least near-canonical) form, using the symmetry properties of the projector itself; note that the projector has the same symmetry properties as the product of transverse metric tensors to which it is dual. For our example the canonicalised form would be $P_\perp^{p_1p_1p_2p_4}$. Subsequently the program writes this as \lstinline|Prt4(1,1,2,3,p1,p2,p4)|. Here the integers in the contraction pattern $1123$ refer to the position of momenta in the second argument list $p_1,p_2,p_4$. The upshot is that although the integral may have 4 loop momenta $p_1,\dots,p_4$ only 3 of those actually appear in the integrand. The result of the contraction \lstinline|Prt4(1,1,2,3,p1,p2,p4)| is then evaluated through \FORM{}'s more recent id-table structure with 1123 denoting the table element which itself is defined as a function of three momenta \lstinline|p1,p2,p3|, which in the substitution are replaced respectively with \lstinline|p1,p2,p4|. So, for example, the corresponding table element is defined in \OPITER{} as follows:
\begin{lstlisting}
Table sparse, Prt4(4,p1?,p2?,p3?);

Fill Prt4(1,1,2,3) = dt(p1,p1)*dt(p2,p3)
                     *rat(Dt+1,Dt^3+Dt^2-2*Dt)
                   + dt(p1,p2)*dt(p1,p3)
                     *rat(-2,Dt^3+Dt^2-2*Dt);
\end{lstlisting}
There are substantial advantages for evaluating contracted projectors with this procedure. The first is that the number of different contractions is reduced to a minimal set. 
The second is due to simplifications of the contracted projectors in comparison to the uncontracted projectors which have in general $(n-1)!!$ terms at rank $n$. The contracted projectors tend to have far fewer terms at least when the loop number is less than the rank. An extreme example is given by \lstinline|Prt20(1,...,1,p1)|. The number of terms in the rank 20 projector is 654,729,075. After contraction with 20 identical momenta only 1 term remains.
The tabulisation can thus save vast amounts of algebra during the tensor reduction. In \OPITER{} we have included id-tables containing all 4-loop contractions up to rank 12, all 3-loop contractions up to rank 16, and all 2-loop contractions up to rank 20. Contraction patterns which are not tabulated in this manner are evaluated by explicit contraction with the general projector in the symmetric basis from ref.~\cite{Goode:2024mci}. In this way \OPITER{} can be used at any loop order, but the performance will be affected for higher rank contractions, which are done on the fly. In principle, the tables could be extended to higher rank and loop numbers, at the cost of larger table files and thus also longer loading times. The authors can supply extended tables upon request by email.

\section{Performance tests and checks}
\label{sec:checks}
To check the output of the code we performed a number of cross-checks. The first of these was to calculate Gaussian-like integrals of the form 
\begin{equation}\label{eq:check_intergral}
    I^{\ixset{\mu}{n}}_E = \int \dd^Dp\, p^{\mu_1}\dots p^{\mu_n}e^{-p^2+2p\cdot(q_1 +\dots+ q_E)}
\end{equation} 
These integrals can be directly computed from standard Gaussian results 
\begin{equation}
    I^{\ixset{\mu}{n}}_E =\frac{1}{2^{n+1}}\Omega(D)\Gamma(D/2) \frac{\partial}{\partial q_1^{\mu_1}}\dots \frac{\partial}{\partial q_1^{\mu_n}} e^{(q_1 +\dots +q_E)^2},
\end{equation}
where $\Omega(D) = 2\pi^{D/2}/\,\Gamma(D/2)$ is the surface area of a D-dimensional sphere and $\Gamma$ is the usual Euler-Gamma function. Inputting the integrand of \cref{eq:check_intergral} in to \OPITER{} results in scalar integrals of the form  
\begin{equation}
    \begin{split}
    I_{E,n,m_1,\dots,m_E} = \int \dd^Dp\, (p^{2})^n(p\cdot q_1)^{m_1}\dots (p\cdot q_1)^{m_E}  e^{-p^2+2p\cdot(q_1 +\dots+ q_E)}\\
    = \frac{(-1^n)\Omega(D)\Gamma(D/2)}{2^{m_1+\dots+m_E+1}}\frac{\partial^n}{\partial a^n}\frac{\partial^{m_1}}{\partial b_1^{m_1}}\dots \frac{\partial^{m_E}}{\partial b_E^{m_E}}
\frac{e^{(b_1q_1 +\dots +b_Eq_E)^2/a}}{a^{D/2}}\Big|_{\substack{a=1\\b_i=1}}. \nonumber
    \end{split}
\end{equation}

To test multi loop examples we may simply multiply several integrals of the form in \cref{eq:check_intergral}. \OPITER{} will  reduce all the loop momenta at once so this will be a valid test of the method. The output of \OPITER{} will now include integrals of the form
\begin{equation}
    \begin{split}
    \int \dd^Dp_1\,&(p_1.p_2)^{r_2}\dots (p_1.p_L)^{r_{L}} (p_1^{2})^n \\ &\times(p_1\cdot q_1)^{m_1}\dots (p_1\cdot q_1)^{m_E} e^{-p_1^2+2p_1\cdot(q_1 +\dots+ q_E)}
    \end{split}
\end{equation}
We then factorise out the other loop momenta and  solve integrals of the general form 
\begin{align}
    & \int \dd^Dp\, p^{\mu_1}\dots p^{\mu_r}(p^{2})^n(p\cdot q_1)^{m_1}\dots (p\cdot q_1)^{m_E}  e^{-p^2+2p\cdot(q_1 +\dots+ q_E)}\nonumber\\
    &= \frac{(-1^n)\Omega(D)\Gamma(D/2)}{2^{m_1+\dots+m_E+r+1}}\frac{\partial^r}{\partial Q^{\mu_1}\cdots \partial Q^{\mu_r} }\frac{\partial^n}{\partial a^n} \frac{\partial^{m_1}}{\partial b_1^{m_1}}\dots \frac{\partial^{m_E}}{\partial b_E^{m_E}} \frac{e^{\frac{\hat Q^2}{a}}}{a^{D/2}}\Big|_{\substack{a=1\\b_i=1\\Q=0}}\,.\nonumber
\end{align}
where $\hat Q=b_1q_1 +\dots +b_Eq_E+Q$.

This represents a very non-trivial check of our code as many exact cancellations need to occur for the denominators to cancel at the end of the calculation. Any small error in terms of signs, coefficients or combinatorics would result in an incorrect answer. We performed this cross-check up to 4 external momenta, 3 loops and up to 8 Lorentz indices.

Further, the projectors where thoroughly checked in ref. \cite{Goode:2024mci} and projectors up to rank $\sim 14$ have been applied in calculations of physically meaningful quantities in the context of the $R^*$-method \cite{Herzog:2017bjx,Herzog:2017dtz,Herzog:2018kwj,Herzog:2017ohr}.

The consistency of the transformation to the van Neerven-Vermaseren basis has been checked by forward and backwards transformation up to 8 external momenta.

\subsection{Performance check}
\begin{figure}
    \centering
    \includegraphics[width=0.45\textwidth]{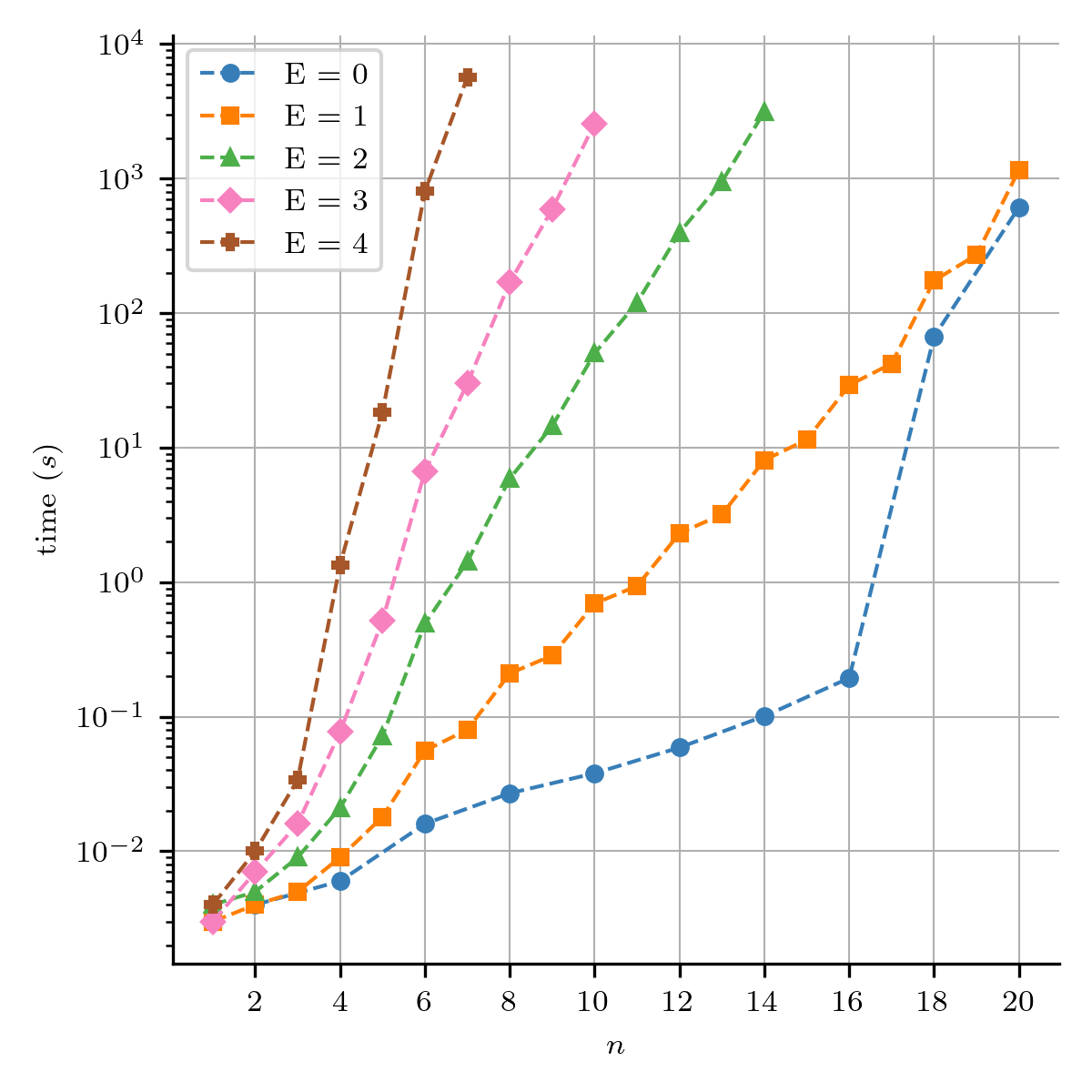}
    \caption{The runtime (in seconds) of \OPITER{} plotted against tensor rank $n$ for the family of integrands $T(n)= k_1^{\mu_1}k_1^{\mu_2}k_2^{\mu_3}k_2^{\mu_4}k_3^{\mu_5}\dots k_3^{\mu_n}$ for several values of $E$. For the case of $E=0$ the odd values of $n$ vanish and have been omitted. For $n<5$ the expression for $T(n)$ is truncated at that tensor rank so these low rank examples are no longer 3-loop.}
    \label{fig:benchmarks}
\end{figure}

We will now consider some sample tensor integrals to showcase the performance of \OPITER.
For this purpose we consider the family of 3-loop Feynman tensor numerators
\begin{equation}
\label{eq:Tn}
T(n)= k_1^{\mu_1}k_1^{\mu_2}k_2^{\mu_3}k_2^{\mu_4}k_3^{\mu_5}\dots k_3^{\mu_n}\,.
\end{equation}
We time the running of the program for various values of $E$. We use \lstinline|tensormode = 2| and \lstinline|tensorbasis = 1| and subtract the time taken to read in all the tables as this will always be performed exactly once at the beginning of the program no matter how many terms are reduced. The timings are presented in \cref{fig:benchmarks}. It is apparent that the growth approximately follows a power law, though increasing $E$ raises the run time by roughly an order of magnitude for a given $n$. The rapid jump in runtime for $E=0$ after $n=16$ is explained by the program switching from the tablelised projectors to contracting the full projector on the fly. For other values of $E$ this effect is obscured by the effects of more external momenta.

\section{Conclusion and Outlook}
\label{sec:conclusion}
In this article we introduced the \OPITER{} program, a procedure for tensor reduction of multi-loop tensor Feynman integrals with tensorial rank up to 20, and
depending on up to 8 external momenta. \OPITER{} works by splitting tensor integrals into transverse and parallel components which is achieved via the van Neerven-Vermaseren basis. 
The transverse parts are subsequently reduced using projectors previously derived by the authors in ref. \cite{Goode:2024mci} via an orbit partition approach. A further feature implemented in \OPITER{} is that it makes use of a basis of tensors which is invariant under integrand symmetries due to exchanges of Lorentz indices. This in effect allows \OPITER{} to tame the factorial growth which is usually encountered with increasing tensor rank. \OPITER{} is also able to deal with tensor integrals with spinor indices in a fully $D$-dimensional setting. This is achieved efficiently by passing into the antisymmetric basis of gamma matrices.

While \OPITER{} is a multi-purpose tensor reduction tool applicable for arbitrary covariant (pseudo-)Euclidean Tensor integrals, we envision that it will be particularly useful for calculations in the context of renormalisation and/or asymptotic expansions, where differential operators are employed to create high-rank tensors. However \OPITER{} could also become useful for calculations involving non-standard tensor integrals, as they may appear, for example, in cosmology \cite{Lee:2024nqu}. In another vein \OPITER's transverse decomposition features could be useful also in the context of IBP reduction which make use of this decomposition \cite{Mastrolia:2016czu, Abreu:2017xsl, Chestnov:2024mnw}.

The performance of \OPITER{} is particularly good for vacuum integrals and slows down in the presence of more external momenta. To improve this it could be useful to implement the Wick-contraction formula of ref.\ \cite{Anastasiou:2023koq} in some future upgrade.

\section*{Acknowledgments}
FH is supported by the UKRI FLF ``Forest Formulas for the LHC'' Mr/S03479x/1. JG,
FH and ST are supported by the STFC Consolidated Grant ``Particle Physics at the
Higgs Centre''. FH would like to thank Jos Vermaseren for implementing id-tables into \FORM, a command which was developed specifically for this project.

\appendix

\section{Derivation of the $c(\alpha)$ factor} \label{sec:caproof}
We present a proof of the expression for the combinatorial factor $c(\alpha)$ appearing in \cref{eq:dperpcontracted}. The contraction of the tensorial part of the integrand with $d_\perp$ contracts the indices in all possible ways.  To  find $c(\alpha)$, we must count how many ways there are to construct the monomial  
\begin{equation}
    m(\alpha)=\prod_{i\le j}  (p_i.p_j)^{\alpha_{ij}}\,,
\end{equation}
which is specified by the matrix $\alpha$. $\alpha$ has the important properties:
\begin{equation} \label{eq:proofalphacon}
      \alpha_{ij}=\alpha_{ji}\,, \qquad \sum_{j=1}^L \alpha_{ij} + \alpha_{ii} = N_i\,.
\end{equation}
We begin by counting the contribution from the diagonal elements. Each momentum $p_i$ has multiplicity $N_i$. The term $p_i\cdot p_i$ appears $a_{ii}$ times, so the number of ways of doing these parings is 
\begin{equation}
    \frac{1}{\alpha_{ii}!}\binom{N_i}{2}\binom{N_i-2}{2}\cdots \binom{N_i-2\alpha_{ii}}{2} = \frac{\left(N_i\right)_{2\alpha_{ii}}}{\,2^{\alpha_{ii}}\,{\alpha_{ii}}!}\,,
\end{equation}  
where each binomial coefficient $\binom{n}{m}$ correspond to the choosing of two $p_i$ to pair up from the ones that remain, and $(n)_a= \frac{n!}{(n-a)!}$ is the Pochhammer symbol for the falling factorial. The factor of $\frac{1}{\alpha_{ii}!}$ fixes the overcounting from ordering these parings.

We will now work out the contribution from the off-diagonal elements. In this we need only consider the contribution of the upper-triangular elements. For ease of calculation we shall work left to right and top to bottom across these elements. For an off diagonal element $\alpha_{ij}$ we must find the number of ways to pair the remaining $p_i$ and $p_j$ such that we make $\alpha_{ij}$ pairs. The number of ways to chose the $p_j$s will depend on how many $p_j$ remain. The number remaining is given by
\begin{equation}
    N_j - 2\alpha_{jj} - \sum_{k=1,\,k\neq j}^{i-1}\alpha_{kj} = \sum_{k=i,\, k\neq j}^L\alpha_{kj} = n_{i,j}
\end{equation}      
where we subtract the number of $p_j$ used in the diagonal elements and in the elements of $\alpha$ above $\alpha_{ij}$. The first equality is achieved by applying the second constraint in \cref{eq:proofalphacon}. The number of ways of choosing the $\alpha_{ij}$ momenta needed is then 
\begin{equation}
    \left(n_{i,j}\right)\left(n_{i,j}-1\right)\cdots \left(n_{i,j}-\alpha_{ij}+1\right) = \left(n_{i,j}\right)_{\alpha_{ij}}\,.
\end{equation} 
Similarly we may count the remaining $p_i$  as
\begin{equation}
    N_i - 2\alpha_{ii} - \sum_{k=i,\,k\neq i}^{j-1}\alpha_{ik} = \sum_{k=j,\, k\neq i}^L\alpha_{ik} = {n}_{j,i} \,,
\end{equation}  
and so the number of ways to choose them is just $\left({n}_{j,i}\right)_{\alpha_{ij}}$.
Combining these factors, the contribution from an upper-triangular off-diagonal element is given by 
\begin{equation}
    \frac{1}{\alpha_{ij}!} \left(n_{i,j}\right)_{\alpha_{ij}}\,\left({n}_{j,i}\right)_{\alpha_{ij}}\,,
\end{equation}    
where the $\frac{1}{\alpha_{ij}!}$ cancels the overcounting from ordering the pairings.
To find the overall $c(\alpha)$ we must simply take a product over all the upper-triangular elements.
We arrive at the expression:
\begin{equation}
    \begin{split}
    c(\alpha) =& \prod_{i=1}^L \prod_{i<j\leq L} \frac{\Big(N_{i}\Big)_{2\alpha_{ii}}\left(n_{i,j}\right)_{\alpha_{ij}}\,\left({n}_{j,i}\right)_{\alpha_{ij}}}{\,2^{\alpha_{ii}}\,{\alpha_{ii}}!\,\alpha_{ij}!}\,.
    \end{split}
\end{equation}

\section{$H$-invariants}
In this appendix we prove some statements given in section \ref{sec:intSym} about the stabiliser group $H$.
\subsection{H-invariance of $m(\alpha)$} \label{sec:Hinvarc}
To prove that the monomial $m(\alpha)$ is $H$-invariant we will show that there is a bijection between the monomials and orbits of the $t_\perp$ basis under $H$. We first 
define
\begin{equation}
 p^{\mu_1\dots \mu_N}(\sigma) = p_1^{\mu_{\sigma(1)}}\dots p_1^{\mu_{\sigma(N_1)}}p_2^{\mu_{\sigma(N_1+1)}} \dots p_1^{\mu_{\sigma(N_1 +N_2)}} p_L^{\mu_{\sigma(N-N_L+1)}}\dots p_L^{\mu_{\sigma(N)}}\,
\end{equation}
such that 
\begin{equation}
    p(\tau\circ h) = p(\tau)\,, \qquad \forall h\in H\,.
\end{equation}
To prove that two elements in a given orbit share a monomial, consider a (transverse) contraction
\begin{equation}
    t_\perp(\sigma)\cdot p(\tau)=m\, ,
\end{equation}
and now permute $t_\perp(\sigma)$ by an $h\in H$:
\begin{align}
    t_\perp(\sigma\circ h)\cdot p(\tau)=t_\perp(\sigma)\cdot p(\tau\circ h^{-1})=  t_\perp(\sigma)\cdot p(\tau)=m\,,
\end{align}
where we were able to move the $h$ across the ``$\cdot$'' as it represents a contraction of all indices.

Now we will prove that two elements with the same monomial must be in the same orbit. Consider two permutations $\sigma$ and $\sigma'=\sigma \circ g$ for some $g\in S_2^N$ such that 
\begin{equation}
    t_\perp(\sigma)\cdot p(\tau)=m=t_\perp(\sigma')\cdot p(\tau)\,.
\end{equation}
We are free to act on both sides of the ``$\cdot$'' so we do so with $g^{-1}$:
\begin{align}
   m=t_\perp(\sigma'\circ g^{-1})\cdot p(\tau\circ g^{-1}) =t_\perp(\sigma)\cdot p(\tau\circ g^{-1})\,.
\end{align}
$p(\tau)$ must then be invariant under the action of $g^{-1}$ and so $g\in H$. We conclude that $t_\perp(\sigma)$ and $t_\perp(\sigma')$ are in the same orbit.
If two $t_\perp$ have the same $m$ they are in the same orbit and, conversely, every member of an orbit has the same $m$.

\subsection{Form of $T_\perp(\alpha)$} \label{sec:formT}
We will now show that $T_\perp(\alpha)$ has the form presented in \cref{eq:Tdiffrep}.
Since $m(\alpha)$ is a monomial of degree $N$ with degree $N_i$ in each $p_i$ we have that
\begin{equation}
 m(\alpha)=
   \frac{\prod_{i=1}^L  p_i^{\mu_{i,1}}}{N_1!\cdots N_L!}\;\Bigg[
\left(\prod_{i=1}^L \frac{\partial^{N_i} }{\partial p_i^{\mu_{i,1}}\cdots \partial p_i^{\mu_{i,N_i}}}\right)\, m(\alpha)\Bigg|_{p_i=0}\Bigg].
\end{equation}
The term in the square brackets is clearly $H$-invariant, which means that it must contain a sum over all $t_\perp$ in a particular orbit under the action of $H$.
It can not contain several orbits since the contraction with the loop momenta yields back $m(\alpha)$ which characterises a particular orbit and all the numerical coefficients generated by the differential operator are positive, meaning terms can not cancel when contracted. Therefore the term is proportional to $T_\perp^{\vec{\mu}}(\alpha)$. The constant of proportionality is determined by demanding that
\begin{equation}
m(\alpha)c(\alpha)=T_\perp^{\,\overbrace{p_1\dots p_1}^{N_1}  \overbrace{p_2\dots p_2}^{N_2}\dots \overbrace{p_L\dots p_L}^{N_L}}
\end{equation}
since there are $c(\alpha)$ elements in the orbit who all contract to the same monomial $m(\alpha)$. From this the result of \cref{eq:Tdiffrep} follows.

\section{Step-by-step output of procedures} \label{sec:procexample}
In this appendix we demonstrate the output of each of the main procedures that make up \OPITER. We have provided the file \lstinline|proc_example.frm| that prints the output of each of highest level procedures acting on a simple example. It may be run from the terminal with 
\begin{lstlisting}
form proc_example.frm
\end{lstlisting}
The example that is expanded is
\begin{lstlisting}
L [without_sym] = ext(q1)*loop(p1,p2)*p1(mu1)
                  *p2(mu2)*p2(mu3);
L [with_sym] =    ext(q1)*loop(p1,p2)*p1(mu1)
                  *p2(mu2)*p2(mu3);   
\end{lstlisting}
To the first expression we will apply the procedures selected when the setting \lstinline|#define tensormode "1"| is chosen and for the second expression we will use integrand symmetry. To expand tensors without integrand symmetry we invoke the \lstinline|tenred| procedure:
\begin{lstlisting}
#call tenred
\end{lstlisting}
This has the following output
(where the \lstinline|ext()| and \lstinline|loop()| functions have been suppressed):
\begin{lstlisting}
[without_sym] =
    + ddts(mu1,mu2)*dual(q1,mu3) * 
      dt(p1,p2)*p2.q1*rat(1,D - 1)
    + ddts(mu1,mu3)*dual(q1,mu2) * 
      dt(p1,p2)*p2.q1*rat(1,D - 1)
    + ddts(mu2,mu3)*dual(q1,mu1) * 
      dt(p2,p2)*p1.q1*rat(1,D - 1)
    + dual(q1,mu1)*dual(q1,mu2)*dual(q1,mu3) *
      p1.q1*p2.q1^2*rat(1,1);
\end{lstlisting}
Similarly 
\begin{lstlisting}
#call tenredisym
\end{lstlisting}
has the following output:
\begin{lstlisting}
[with_sym] =
    + sym(ind1(mu2,mu3)*ind2(MMu2,MMu3))*
      ddts(MMu2,MMu3)*dual(q1,mu1) * ( 
      dt(p2,p2)*p1.q1*rat(1,D - 1) )
    + sym(ind1(mu2,mu3)*ind2(MMu2,MMu3))*
      ddts(MMu2,mu1)*dual(q1,MMu3) * ( 
      dt(p1,p2)*p2.q1*rat(2,D - 1) )
    + sym(ind1(mu2,mu3)*ind2(MMu2,MMu3))*
      dual(q1,MMu2)*dual(q1,MMu3)*dual(q1,mu1)* 
      ( p1.q1*p2.q1^2*rat(1,1) );
\end{lstlisting}
For a description of the \lstinline|sym| function, see \cref{sec:symmetrizer}. While for higher rank examples the use of the \lstinline|sym| function can lead to drastic size reduction of the output,the gain is not significant for this simpler example. In the following we will demonstrate the effect of the various procedures on the \lstinline|[without_sym]| expression. 

We now expand dot products of transverse momenta \lstinline|dt(p,p)| in terms of dot products of momenta and inverse-Gram matrix elements (these are denoted \lstinline|H(i,j)|) with 
\begin{lstlisting}
#call expanddt
\end{lstlisting} 
This results in the following:
\begin{lstlisting}
[without_sym] =
    + ddts(mu1,mu2)*dual(q1,mu3) * 
        (p1.p2*p2.q1*rat(1,D - 1) 
          + H(1,1)*p1.q1*p2.q1^2*rat(-1,D - 1))
    + ddts(mu1,mu3)*dual(q1,mu2) * 
        (p1.p2*p2.q1*rat(1,D - 1) 
          + H(1,1)*p1.q1*p2.q1^2*rat(-1,D - 1))
    + ddts(mu2,mu3)*dual(q1,mu1) * 
        (p1.q1*p2.p2*rat(1,D - 1) 
          + H(1,1)*p1.q1*p2.q1^2*rat(-1,D - 1))
    + dual(q1,mu1)*dual(q1,mu2)*dual(q1,mu3) * 
        ( p1.q1*p2.q1^2*rat(1,1) );
    \end{lstlisting} 
In this case the inverse-Gram element is particularly simple as there is only one external momenta so \lstinline|H(1,1)|$=1/q_1^2$
When choosing the setting \lstinline|#define tensorbasis "2"|, \OPITER{} will move out of the dual-transverse basis after the tensor reduction has been performed.
The first step of the transformation is expanding these transverse metrics \lstinline|ddts(mu1,mu2)| with 
\begin{lstlisting}
id ddts(mu1?,mu2?)=dt(mu1,mu2);
#call expanddt
\end{lstlisting}
The result is:
\begin{lstlisting}
[without_sym] =
    + dual(q1,mu1)*d_(mu2,mu3) * 
        (p1.q1*p2.p2*rat(1,D- 1) 
         + H(1,1)*p1.q1*p2.q1^2*rat(-1,D - 1))
    + dual(q1,mu1)*dual(q1,mu2)*dual(q1,mu3) * 
        (p1.q1*p2.q1^2*rat(1,1))
    + dual(q1,mu1) * 
        (H(1,1)*q1(mu2)*q1(mu3)*
            p1.q1*p2.p2*rat(-1,D - 1) 
         + H(1,1)^2*q1(mu2)*q1(mu3)*
            p1.q1*p2.q1^2*rat(1,D - 1))
    + dual(q1,mu2)*d_(mu1,mu3) * 
        (p1.p2*p2.q1*rat(1,D- 1) 
        + H(1,1)*p1.q1*p2.q1^2*rat(-1,D - 1))
    + dual(q1,mu2) * 
        (H(1,1)*q1(mu1)*q1(mu3)*
            p1.p2*p2.q1*rat(-1,D - 1) 
        + H(1,1)^2*q1(mu1)*q1(mu3)*
            p1.q1*p2.q1^2*rat(1,D - 1))
    + dual(q1,mu3)*d_(mu1,mu2) * 
        (p1.p2*p2.q1*rat(1,D- 1) 
         + H(1,1)*p1.q1*p2.q1^2*rat(-1,D - 1))
    + dual(q1,mu3) * 
        (H(1,1)*q1(mu1)*q1(mu2)*
            p1.p2*p2.q1*rat(-1,D - 1) 
         + H(1,1)^2*q1(mu1)*q1(mu2)*
            p1.q1*p2.q1^2*rat(1,D - 1));
        \end{lstlisting}
All the \lstinline|ddts| have been expanded now though inverse-Gram elements and \lstinline|dual|s remain.
The next step is to expand the dual momenta which is done by calling
\begin{lstlisting}
#call dual2ext
\end{lstlisting}
The result is:
\begin{lstlisting}
    [without_sym] =
    + q1(mu1)*q1(mu2)*q1(mu3)*p1.q1*p2.q1^2 * 
        (H(1,1)^3*rat(D + 2,D - 1))
    + q1(mu1)*q1(mu2)*q1(mu3)*p1.q1 *
        (H(1,1)^2*p2.p2*rat(-1,D - 1))
    + q1(mu1)*q1(mu2)*q1(mu3)*p2.q1 *
        (H(1,1)^2*p1.p2*rat(-2,D - 1))
    + d_(mu1,mu2)*q1(mu3)*p1.q1*p2.q1^2 * 
        (H(1,1)^2*rat(-1,D - 1))
    + d_(mu1,mu2)*q1(mu3)*p2.q1 * 
        (H(1,1)*p1.p2*rat(1,D - 1))
    + d_(mu1,mu3)*q1(mu2)*p1.q1*p2.q1^2 * 
        (H(1,1)^2*rat(-1,D - 1))
    + d_(mu1,mu3)*q1(mu2)*p2.q1 * 
        (H(1,1)*p1.p2*rat(1,D - 1))
    + d_(mu2,mu3)*q1(mu1)*p1.q1*p2.q1^2 * 
        (H(1,1)^2*rat(-1,D - 1))
    + d_(mu2,mu3)*q1(mu1)*p1.q1 * 
        (H(1,1)*p2.p2*rat(1,D - 1));
    \end{lstlisting}

Finally, we substitute the elements of the inverse Gram matrix as well as its determinant.
This is done by calling the following two procedures:
\begin{lstlisting}
#call subinvgram
#call subinvgramdet
\end{lstlisting} 
This brings us to the final result of the reduction:
\begin{lstlisting}
[without_sym] =
    + q1(mu1)*q1(mu2)*q1(mu3)*
        p1.p2*p2.q1*q1.q1^-2*rat(-2,D - 1)
    + q1(mu1)*q1(mu2)*q1(mu3)*
        p1.q1*p2.p2*q1.q1^-2*rat(-1,D - 1)
    + q1(mu1)*q1(mu2)*q1(mu3)*
        p1.q1*p2.q1^2*q1.q1^-3*rat(D + 2,D - 1)
    + d_(mu1,mu2)*q1(mu3)*
        p1.p2*p2.q1*q1.q1^-1*rat(1,D - 1)
    + d_(mu1,mu2)*q1(mu3)*
        p1.q1*p2.q1^2*q1.q1^-2*rat(-1,D - 1)
    + d_(mu1,mu3)*q1(mu2)*
        p1.p2*p2.q1*q1.q1^-1*rat(1,D - 1)
    + d_(mu1,mu3)*q1(mu2)*
        p1.q1*p2.q1^2*q1.q1^-2*rat(-1,D - 1)
    + d_(mu2,mu3)*q1(mu1)*
        p1.q1*p2.p2*q1.q1^-1*rat(1,D - 1)
    + d_(mu2,mu3)*q1(mu1)*
        p1.q1*p2.q1^2*q1.q1^-2*rat(-1,D - 1);
    \end{lstlisting}
The tensor has now been fully reduced and Lorentz indices appear only on the originally defined external momenta and full metrics represented by \FORM's inbuilt \lstinline|d_|.


\bibliographystyle{JHEP}
\renewcommand*{\bibfont}{\justify}
\bibliography{refs.bib}







\end{document}